# Optical Determination of Electron-Phonon Coupling in Carbon Nanotubes

*Y. Yin,[1] A. Vamivakas,[2] A. Walsh,[1] S. Cronin,[3] M. S. Ünlü,[2,1] B. B Goldberg,[1,2] and A. K. Swan[2]*

[1]Physics Department, [2]Electrical and Computer Engineering Department, Boston University, Boston, MA 02215, [3]Electrical Engineering Department, University of Southern California, Los Angeles, CA 90089

We report on an optical method to directly measure electron-phonon coupling in carbon nanotubes by correlating the first and second harmonic of the resonant Raman excitation profile. The method is applicable to 1D and 0D systems and is not limited to materials that exhibit photoluminescence. Experimental results for electron-phonon coupling with the radial breathing mode in 5 different nanotubes show coupling strengths from 3-11 meV, depending on chirality. The results are in good agreement with the chirality and diameter dependence calculated by Goupalov *et al*.

The intrinsic coupling between the electronic and atomic degrees of freedom is fundamental to many transport and optical properties of condensed matter. Here we examine carbon nanotubes, a prototypical one-dimensional (1D) system under intense study for both fundamental physical properties and potential applications as transistors, sensors and optoelectronic devices. The electron-phonon (e-ph) coupling in carbon nanotubes (CNTs) is a key coupling parameter in the electronic band structure, controls Raman scattering [1], heat and electron conductivity [2], and sets an upper limit to ballistic transport [3, 4]. In polar semiconductors, multi-phonon replicas in luminescence have been used to measure the Fröhlich e-ph coupling with longitudinal optical (LO) phonons in good agreement with theory [5, 6]. In low dimensional systems, the electron and phonon confinement complicates the situation, with conflicting experimental results [7-10], and requiring model-dependent interpretation to extract the e-ph interaction. E-ph interactions in CNTs are due to the deformation-potential interaction, generally weaker than Fröhlich interaction in polar materials. Nonetheless, evidence of e-ph coupling in CNTs are observed in photoluminescence phonon replicas [11], the Kohn anomaly manifested in the phonon dispersion and line-widths in metallic nanotubes [12] and temperature dependence of optical transition energies [13, 14]. However, direct measurements of e-ph coupling in CNTs are lacking.

Recent tight-binding calculations predicts a strong chirality and family dependence of the e-ph coupling of the radial breathing mode (RBM) in carbon nanotubes [15, 16]. The relative strengths of the e-ph couplings are supported by Raman intensity measurements from a CNT ensemble, assuming a uniform chirality distribution [15]. Generally, both optical and transport ensemble measurements are challenged by the variation of electronic and phonon structure with chirality, and advancements in single tube measurements have been made with knowledge of the *(n,m)* structure [2, 17-19]. In this letter, we extract an absolute value of the RBM e-ph coupling for singly resonant, suspended nanotubes of 5 different chiralities that accurately follow the chiral dependence suggested from calculations [15]. Our general approach is enabled by using the full resonant Raman profile, thus simply accounting for contributions from the optical matrix elements, resonant enhancement factors, and collection and detection efficiencies.

We investigate a series of singly optically resonant single wall carbon nanotubes (SWNTs) in small ropes with different diameters and chiralities, grown by chemical vapor deposition over 1-1.5μm wide trenches etched in quartz substrates [13]. Fig. 1(a) shows a scanning electron microscopy (SEM) picture of a sample. Raman excitation (Ti:S 720-830 nm) with tilt-tuning of matching filters and use of a fast single grating monochrometer offer a high throughput system enabling single tube detection, even for two-phonon Raman scattering [20]. Powers less than 2mW in a 0.5μm spot size avoids tube heating, and for each excitation energy, all the different Raman modes are collected simultaneously.

Fig. 1(b) and 1(c) show the resonant Raman scattering excitation (RRSE) spectra and profiles of the Stokes, anti-Stokes, and two-phonon Stokes radial breathing mode from a tube suspended in air. Fig. 1(b) shows the raw 2D spectral data map after subtracting a linear background. The anti-Stokes (AS, left) and Stokes (S, middle) resonances are both clearly observed, and their intensity maxima are shifted in excitation energy by the RBM phonon energy due to the resonant enhancement for both the incoming and scattered light [20-22]. The two-phonon RBM peak is much weaker and shifted to higher excitation energy due to the two-phonon resonant enhancement of scattered light. The resulting RRSE profiles from the AS, S RBM and S two-phonon RBM (2RBM) peaks are shown in Fig. 1(c). After fitting the Stokes RBM profile with Eq. (1), we use the same parameters to calculate the anti-Stokes scattering profile, matching the AS data with no adjustable parameters, demonstrating that the tube is not heated (300K ± 15K). The two-phonon RRSE profile is fitted with Eq. (2) and comparison of the profiles from one-phonon and two-phonon scattering is then used to extract the absolute value of e-ph coupling matrix elements.

The resonant Raman cross section for first order, one-phonon scattering is given by [22, 23]

$$\frac{d_1\sigma'(E_L, E_{ph}, \theta)}{d\Omega} = A\frac{(E_L - E_{ph})^2}{E_{ph}^2 E_L^2}\left|\frac{1}{\sqrt{E_L - E_{ii} - i\eta}} - \frac{1}{\sqrt{E_L - E_{ph} - E_{ii} - i\eta}}\right|^2 \quad (1)$$

where we define $A = CN_{ph}|M_{e-ph}|^2$. $C$ is a tube dependent constant which includes photon energy independent parts of the optical matrix elements. In our calculation, only one electron and hole band with a transition energy $E_{ii}$ are considered; $E_L$ is the excitation photon energy; $E_{ph}$ is the phonon energy; $\eta$ is the broadening factor in units of energy. $M_{e\text{-}ph}$ is the e-ph transition matrix element for spontaneous emission, and the overall e-ph coupling strength in a one-phonon RRS process is $N_{ph}|M_{e-ph}|^2$, where $N_{ph} = (n_{ph} + 1)$ for Stokes scattering and $N_{ph} = n_{ph}$ for anti-Stokes scattering; and $n_{ph} = 1/(e^{|E_{ph}|/kT} - 1)$ is the phonon number. Only zone center (Γ point) phonons are involved in one-phonon first order Raman processes, so $M_{e\text{-}ph}$ is the e-ph coupling at the Γ point. Eq. (1) is used to fit the Stokes resonant excitation profiles, obtain the values of $A$, $E_{ii}$ and $\eta$ for a particular phonon mode of a particular SWNT and calculate the anti-Stokes resonant Raman profiles.

Two-phonon scattering is more complex and includes both simultaneous and sequential emission. For simultaneous emission of two phonons at the Γ point, the RRSE profile has two resonant peaks at $E_{ii}$ and $E_{ii}+2E_{ph}$ [23]. For sequential emission of the two-phonons at Γ point, RRSE profile has three resonant peaks at $E_{ii}$, $E_{ii}+E_{ph}$ and $E_{ii}+2E_{ph}$, in which the middle peak at $E_{ii}+E_{ph}$ carries twice the weight of other two resonances. It is generally believed that simultaneous emission is negligible compared to sequential emission of two-phonons under normal phonon densities [24], and the strong peaks at $E_{ii}+E_{ph}$ in our measured two-phonon RRSE profiles support this assumption. In our calculation, we only consider the sequential emission of two phonons.

Two-phonon Raman scattering lifts the restriction limiting one-phonon scattering to the Γ point, since two phonons with opposite wave vectors will satisfy the total momentum conservation. This allows non-Γ point phonons to mediate the scattering process and thus makes possible the observation of scattered photon energies that are not simply twice the energies observed in one-phonon Raman scattering. In the measured two-phonon Raman spectra we observe one two-phonon peak at 2xRBM frequency (Fig. 1(b)) indicating a contribution from phonons near the Γ point. To determine the weight of the contribution from the phonons near the Γ point, we need the density of states (DOS) of the particular phonon mode being probed. We calculate the DOS from the phonon dispersion using a microscopic mass and spring model of the nanotube lattice dynamical properties first developed by Mahan and Jeon for achiral [25] tubes and later extended to chiral SWNTs [26]. Fig. 2 shows calculated RBM phonon dispersion curves and the resulting DOS for SWNTs (9,4) and (11,0). Not surprisingly, the DOS has a strong singularity at the Γ point. For the higher symmetry (11,0) tube, we observe an additional singularity at the zone boundary. Using the RBM DOS $\rho_{DOS}^{RBM}(\Omega)$ we determine the weight from the Γ point, $w_\Gamma = \int \rho_{DOS}^{RBM}(\Omega) d\Omega$ where the integration extends over the Γ point singularity, and where the integral of DOS over the entire frequency range is normalized to unity. Our calculation shows that $w_\Gamma \approx 1$ for RBM in chiral SWNTs, while for the achiral (11,0) CNT $w_\Gamma = 0.8$. We now approximate the Γ point phonon contribution to the two-phonon Raman scattering cross-section as:

$$\frac{d_1\sigma'(E_L, E_{ph}, \theta)}{d\Omega} = A' \frac{(E_L - 2E_{ph})^2}{E_L^2} \left| \frac{(p_1 + p_2 + p_3)}{(p_1 + p_3)(p_1 + p_2)(p_3 + p_2)} \cdot \frac{1}{p_1 p_2 p_3} \right|^2 \tag{2}$$

where we define $p_j = \sqrt{E_L - E_{ii} - (j-1)E_{ph} - i\eta}$, $j = 1,2,3$ and $A' = CN_{ph}^2 |M_{e-ph}|^4 w_\Gamma$ [27]. The constant $C$ is the same constant used in the one-phonon scattering, and $w_\Gamma$ is the phonon DOS weighting factor.

The simultaneous measurement of the one-phonon and two-phonon resonant Raman profile spectra for a single mode, such as the RBM, allows us to directly determine the e-ph coupling for the Γ point phonons. We fit the experimental results with Eq. (1) and Eq. (2) to obtain $A$ and $A'$. Then, we can use the equation below to obtain the absolute values of e-ph coupling matrix elements.

$$|M_{e-ph}|^2 = \frac{A'/w_\Gamma}{AN_{ph}} \tag{3}$$

The Stokes RBM profile in Fig. 1(c) is fitted with Eq. (1) to obtain the $A$, $E_{ii}$ (here $E_{ii}$ is $E_{22}$) and $\eta$ values. The assignment of (10,2) is given based on its RBM frequency ($\omega_{RBM}$) and $E_{22}$ transition energy [20, 21]. The SWNT's diameter ($d_t$), chiral angle (θ), and chiral index ν=mod(n-m,3) are also obtained from the assignment. Typically, we are only able to measure about 2/3 of the resonant range of the two-phonon scattering due to our limited excitation tunability. The two-phonon RBM excitation profile for (10,2) is fitted with Eq. (2) and is plotted in Fig. 1(c). In the fitting of two-phonon RRSE profile, the broadening factor $\eta$ is fixed at the value from the corresponding one-phonon scattering. The parameter $E_{ii}$ is allowed to vary independently to account for the approximation in Eq. (2) of only considering phonons strictly at the Γ point. Next, the value of $|M_{e-ph}|^2$ is obtained from Eq. (3) using the calculated $w_\Gamma$ value. Quantitative values for this SWNT are listed in table I. Errors in $|M_{e-ph}|^2$ are calculated from fitting errors of *A* and *A'*, which are assumed to be uncorrelated. RRSE measurements are performed on four additional, different SWNTs: (11,0), (9,4), (9,7) and (14,1). The assignments, profile fitting results and measured RBM $|M_{e-ph}|$ values are listed in table I.

The dependence of RBM e-ph coupling strength is a strong function of the SWNT's diameter, chiral angle and chiral index. Based on their tight binding calculation [15], Goupalov, *et al.* proposed that e-ph coupling for RBM in SWNTs has a form

$$|M_{e-ph}|^2 = \left| \frac{a}{d_t^2} + \nu \frac{b}{d_t} \cos 3\theta \right|^2 \tag{4}$$

where the constants *a* and *b* depend on which electronic energy level $E_{ii}$ is excited. The *b/a* ratio has negative sign for $E_{22}$, resulting in a stronger e-ph coupling for ν= -1. Our experimental data are fit remarkably well ($R^2$=0.996) by Eq. (4), with $a$ = 1.8±0.4 (meV·nm$^2$) and $b$ = −7.7±0.5 (meV·nm) for in-rope SWNTs under $E_{22}$ resonance. Figures 3(a) (ν= −1) and 3(b) (ν=1) plot the coupling strength in a color scale as a function of chiral angle and tube diameter, while Fig. 3(c) shows the same results with all tubes plotted vs. diameter to more accurately display the fitting errors. The errors of $a$ and $b$ are determined by fitting with a 0.95 confidence level. The ratio of $|b/a| = 4.3nm^{-1}$ is slightly larger than the theoretical value $|b/a| = 3/(5a_0) = 4.2nm^{-1}$ [15] using $a_0 = 0.142nm$. The coupling strength (Eq. (4)) has a node for ν=1 shown in Fig. 3(b), which agrees with a prediction made by J. Jiang *et al.* using the tight binding approximation [28]. Later work by the same group [16] using an extended tight binding approximation found similar dependence on diameter and angle, also in agreement with reference [15] and our data. In addition, the measured e-ph coupling strengths can be compared with *ab initio* calculations by M. Machon *et al.* [29]. The *ab initio* e-ph coupling result for the (11,0) tube is larger (28 meV) than our measured value (11±2 meV). We attribute this discrepancy to the difference in e-ph coupling strength between isolated SWNTs and SWNTs in bundles. We will closely examine the difference of bundled and isolated SWNTs in a separate publication.

These results confirm that the e-ph coupling is well described by the deformation potential interaction, derived from the empirical tight binding model. The results demonstrate that the interference of the chirality-dependent and chirality-independent terms indeed lead to stronger interactions for $\nu=-1$ carbon nanotubes at $E_{22}$, in agreement with the many experimental studies that observe stronger Raman cross-sections for $\nu=-1$ for $E_{22}$ transitions. Practically, this means that ensemble Raman measurements must be appropriately scaled to extract the chirality distribution.

In summary, we have proposed and demonstrated an optical method for the direct measurement of e-ph coupling $|M_{e-ph}|$ in carbon nanotubes. By correlating the full resonant excitation profiles of the first and second harmonic Raman peaks of the radial breathing mode and accounting for the phonon density of states, we determined the e-ph coupling strengths for SWNTs under $E_{22}$ resonance for five singly resonant SWNTs in small ropes. The results follow the calculated functional dependence on chiral angle, tube diameter and chiral index $\nu$. This technique does not require luminescence and could be applied to any low dimensional (1D or 0D) system, e.g. nano-rods and quantum-dots, systems which exhibit strong electronic resonance enhancements and well-defined phonon lines.

TABLE I. A list of measured SWNTs with their RBM $|M_{e-ph}|$ values and assignments. For (9,7) SWNT, the $|M_{e-ph}|^2$ value for the RBM is derived from the absolute value of another mode, M273, because the 2RBM is too weak to be measured for (9,7). $|M_{e-ph}|$ values for phonon modes other than RBM will be further discussed in a separate publication.

| | $\omega_{RBM}$ | n,m | $d_t$ (nm) | θ (°) | ν | $E_{22}$ ($eV$) | $\eta$ ($meV$) | $w_\Gamma$ | RBM $A$ | 2RBM $A'$ ($\times 10^{-5}$) | $\lvert M_{e-ph}\rvert^2$ x10(m$eV^2$) | $\lvert M_{e-ph}\rvert$ ($meV$) |
|---|---|---|---|---|---|---|---|---|---|---|---|---|
| 1 | 206.0 | 14,1 | 1.153 | 3.42 | 1 | 1.567 | 14 | 1 | 1.5 | 6.3 | 3±4 | 5±4 |
| 2 | 219.0 | 9,7 | 1.103 | 25.87 | -1 | 1.554 | 38 | 1 | 0.3 | N/A | 0.9±0.6 | 3±1 |
| 3 | 259.0 | 9,4 | 0.916 | 17.48 | -1 | 1.627 | 19 | 1 | 6 | 41 | 5±2 | 7±1 |
| 4 | 260.5 | 10,2 | 0.884 | 8.95 | -1 | 1.591 | 16 | 1 | 1.7 | 24 | 10±4 | 10±2 |
| 5 | 267.0 | 11,0 | 0.873 | 0 | -1 | 1.579 | 13 | 0.8 | 0.9 | 13 | 13±4 | 11±2 |

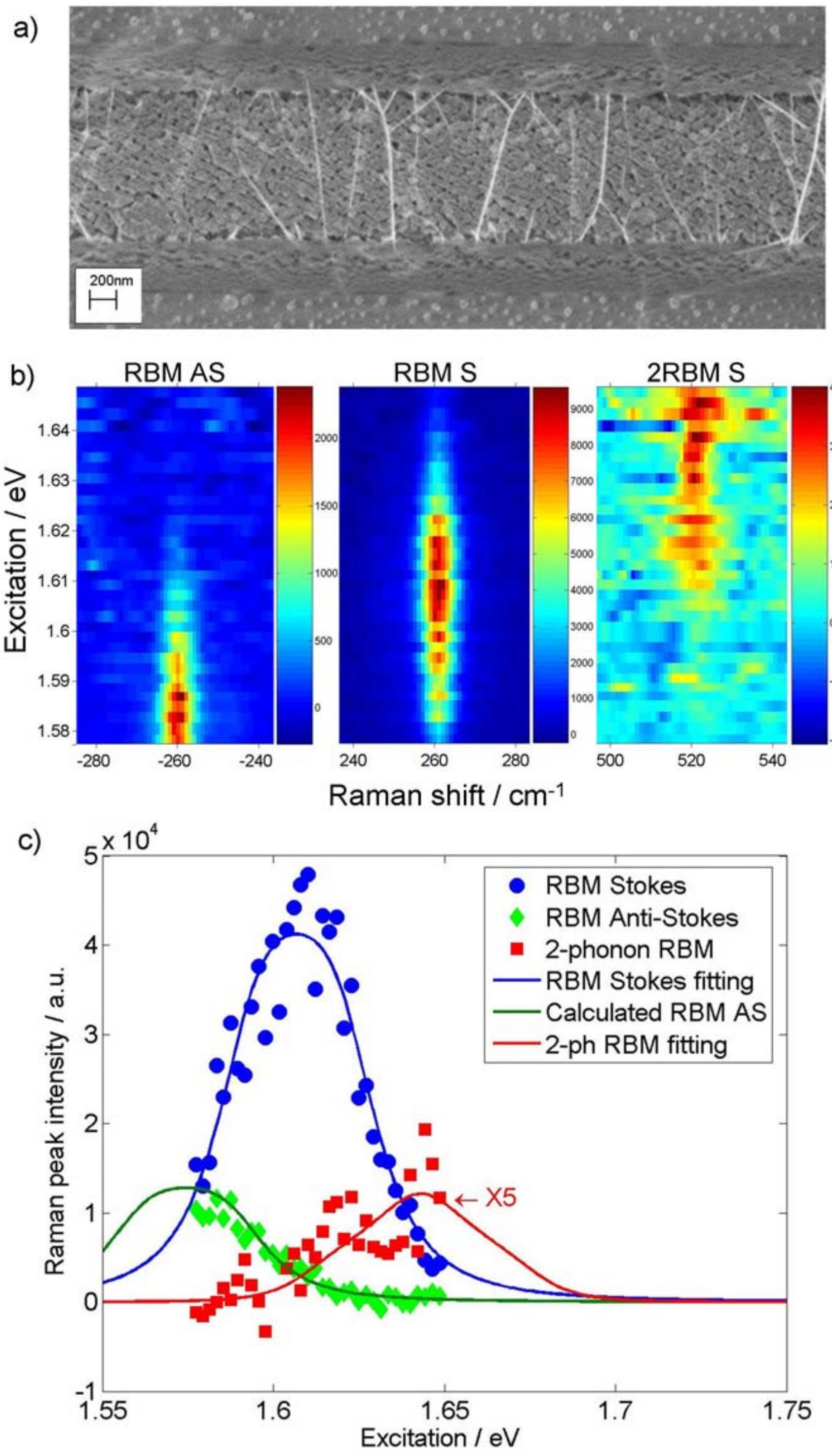

FIG. 1 (color). (a) A SEM picture of a sample showing small, suspended bundles; (b) Resonant Raman scattering excitation spectra of anti-Stokes, Stokes and two-phonon RBM of a singly resonant (10,2) tube; (c) their resonant excitation profiles and fitted or calculated curves. The data and curves for anti-Stokes, Stokes and two-phonon RBM are presented by green (diamond), blue (dot) and red (square) marks and lines. The two-phonon data and curves are magnified by a factor of 5.

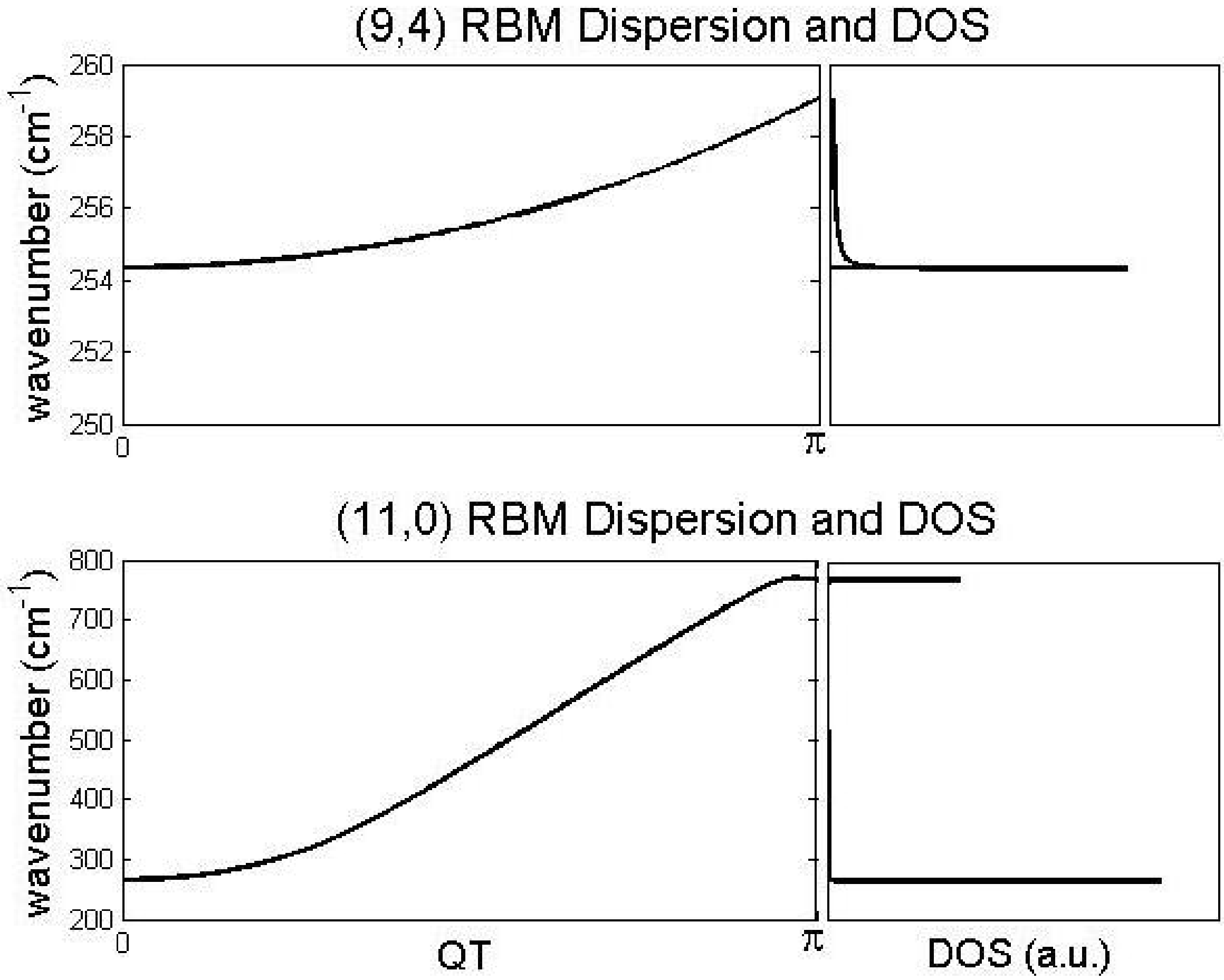

FIG. 2. Calculated RBM phonon dispersion curves and density of states of SWNTs (9,4) and (11,0). QT is the magnitude of the translation vector along the tube axis.

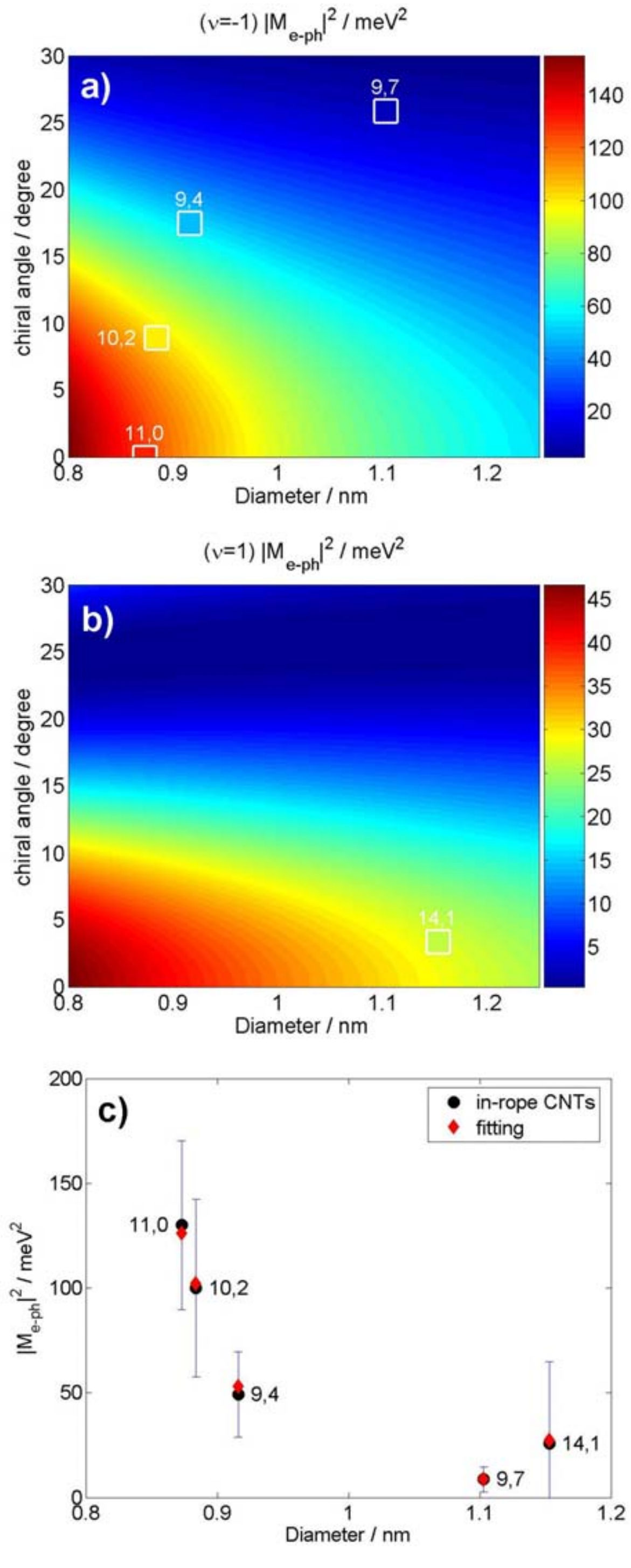

FIG. 3 (color). Measured $|M_{e\text{-}ph}|^2$ values for 5 in-rope SWNTs and the fitting result with Goupalov's formula. (a) and (b) are the 3D color-map plots of $|M_{e\text{-}ph}|^2$ as a function of chiral angle and diameter for SWNTs with chiral index ν=-1 and ν=1; the background colors are the fitted result while measured values are presented as squares with the same color-map scales. (c) plots the $|M_{e\text{-}ph}|^2$ vs. diameter for measured values (black dots) with error bars and fitted results (red diamonds).